\documentclass[12pt,a4paper]{article}
\usepackage{amsmath,amssymb, amsfonts, amsthm}
\usepackage{mathrsfs}
\usepackage{graphicx}

\usepackage{epsfig, euscript}

\usepackage{mathptmx}

\usepackage[shortlabels]{enumitem}
\usepackage{multirow}
\usepackage{longtable}
\usepackage{afterpage}

\usepackage{subfig}
\usepackage{graphicx}
\graphicspath{ {./imgs/} }

\DeclareMathOperator{\Tr}{Tr}

\begin{document}

\title{Modeling  the Eberhard inequality based tests}

\author{Polina Titova\\
Moscow Institute of Electronic Technology\\
124498, Zelenograd, Moscow, Russia Federation\\
Andrei Khrennikov\\
International Center for Mathematical Modeling\\
 in Physics, Engineering, Economics, and Cognitive Science\\
Linnaeus University, S-35195, V\"axj\"o-Kalmar, Sweden\\
}

\maketitle

\begin{abstract} 
Last year the first experimental tests closing the detection
loophole (also referred to as the fair sampling loophole)
were performed by two experimental groups \cite{Zeilinger}, \cite{Kwiat}.
To violate Bell-type inequalities (the Eberhard inequality  in the first 
test and the Clauser-Horne  inequality in the second test), one has 
to optimize a number of parameters involved in the experiment (angles 
of polarization beam splitters and quantum state parameters). Although these are technicalities,
their optimal determination plays an important role in approaching statistically significant 
violations of the inequalities. In this paper we study this problem for the Eberhard inequality in very detail by using 
the advanced method of numerical optimization, the Nelder-Mead method. First of all, we improve
the the results of optimization for the original Eberhard model \cite{Eberhard} and the Gustina et al.  
work \cite{Zeilinger} (``Vienna-13 experiment'') by using the model of this experiment presented in Kofler 
et al. \cite{Zeilinger1}.  We also take into account the well known fact that detectors can have different 
efficiencies and perform the corresponding optimization. 
In previous studies the objective function had the meaning of the mathematical
expectation. However, it is also useful to investigate the possible level of variability
of the results, expressed in terms of standard deviation. In this paper we consider the
optimization of parameters for the Eberhard inequality using coefficient of variation
taking into account possible random fluctuations in the setup of angles during the experiment.
\end{abstract}

\section{Introduction}

Experimental realization of a loophole-free test for Bell \cite{B} inequalities will 
play a crucial role both for quantum foundations \cite{B}, \cite{CH1}-\cite{CH3}, \cite{Aspect}--\cite{Aspect2}, \cite{SH1}, \cite{SH2}   
(see, e.g., \cite{E2}--\cite{KHR_CONT} for recent 
studies) and quantum technologies, e.g., quantum cryptography and quantum random generators. It is clear 
that the often used argument that ``closing of different loopholes in separate tests
can be considered as the solution of the  loopholes problem'' can not be considered as acceptable.
The quantum community put tremendous efforts to perform a loophole-free test  and 
its final realization (which can be expected rather soon) will be  a great event in 
development of quantum theory and quantum technologies. 

Last year the first experimental tests   for photons  closing the {\it detection
loophole} (also referred to as the fair sampling loophole)
were performed by two experimental groups \cite{Zeilinger}, \cite{Kwiat}, see also \cite{Zeilinger2}--\cite{Kwiat1}.
 The detection efficiency problem  
for photons is very complicated and its solution was based on the use of advanced photo-detectors, i.e.,
new technology as well as its testing \cite{POL}. The Bell tests with photons \cite{Aspect}--\cite{Aspect2} 
are promising to close both the detection and locality loopholes, since the latter was closed long ago \cite{Weihs} 
and recently experiments demonstrating 
violation of Bell-type inequalities on large distances  \cite{UTS}--\cite{UTS1} were performed.     
However, to violate Bell type inequalities one has to approach very high efficiency of the total experimental setup.
Hence, although nowadays it is possible to work with photo-detectors having the efficiency approaching 100\%, the loses
in the total experimental setup can decrease essentially the total efficiency of the experimental scheme, see \cite{BO1}-\cite{BO2}
for theoretical analysis and mathematical modeling.  
The experimentalists confront this problem by trying to extend Bell-type tests with sufficiently high efficiency 
to close the locality loophole. The total efficiency of experimental schemes decreases drastically with the distance.
Therefore it is important to optimize all parameters of the experiment to approach the maximal violation for 
the minimal possible efficiency.  (One has 
to optimize angles of polarization beam splitters and the initial state parameters). Although these are technicalities,
they optimal determination play an important role in approaching statistically significant 
violations of the inequalities. 

In this paper we study this problem for the Eberhard inequality \cite{Eberhard} in very detail by using 
the advanced method of numerical optimization, the {\it Nelder-Mead method} \cite{NM}. First of all, we improve
the the results of optimization for the original Eberhard model \cite{Eberhard} and
the Gustina et al.  
work \cite{Zeilinger} (``Vienna-13 experiment'') by using the model of this experiment presented in Kofler 
et al. \cite{Zeilinger1}. We also take into account the well known fact that detectors can have different 
efficiencies and perform the corresponding optimization. 

In the previous studies \cite{Eberhard}, \cite{Zeilinger}, \cite{Zeilinger1} the objective function for parameters optimization had the meaning of the mathematical
expectation. However, it is also useful to investigate the possible level of variability
of the results, expressed in terms of standard deviation. In this paper we consider the
optimization of parameters for the Eberhard inequality using the coefficient $K$ --the  reciprocal of the {\it coefficient of variation}
taking into account possible {\it random fluctuations in the setup of angles during the experiment.}\footnote{In classical signal analysis this quantity 
is known as signal/noise rate \cite{CI1}, \cite{CI2}.}
It seems that our study is the first contribution to this problem, 
This study (of the problem of sensitivity of the degree of violation of the
Eberhard inequality to the precision in the control of angles of polarization beam
splitters) can be useful for experimentalists. One of the results of our numerical simulation 
is unexpected stability of the degree of violation of the Eberhard inequality to fluctuations 
of these angles (in neighborhoods of optimal values of the angles).

\section{Eberhard inequality}
\label{EBB}

We follow Eberhard \cite{Eberhard}: Photons are emitted in pairs $(a,b).$
Under each measurement setting $(\alpha, \beta)$, the events in which the photon $a$ is detected
in the ordinary and extraordinary beams  are denoted by the symbols $(o)$ and $(e)$, respectively, and
the event that it is undetected is denoted by the symbol $(u).$ The same symbols are used to denote the corresponding
events for the photon $b.$ Therefore for the pairs of photons there are nine types of events: $(o, o), (o, u),
(o, e),$ $ (u, o), (u, u), (u, e), (e, o), (e, u),$ and $(e, e).$ 

Under the conditions of locality, realism  and statistical reproducibility the following inequality (the Eberhard inequality)
was derived: 
$$
J\equiv n_{oe}(\alpha_1, \beta_2) +
n_{ou}(\alpha_1, \beta_2)
 + n_{eo}(\alpha_2, \beta_1)+n_{uo}(\alpha_2, \beta_1)+n_{oo}(\alpha_2, \beta_2)
$$
\begin{equation}
\label{FUND}
 - n_{oo}(\alpha_1, \beta_1) \geq 0,
\end{equation}
where $n_{xy}(\alpha_i, \beta_j)$ is the number 
of pairs detected in a given time period for settings $\alpha_i, \beta_j$ with outcomes 
$x,y=o,e, u$ and the outcomes $(o)$ and $(e)$ correspond to detections 
in the ordinary and extraordinary beams, respectively, and
the event that photon is undetected is denoted by the symbol $(u).$  
We point to the main distinguishing features of the E-inequality:

\medskip

a) derivation without the fair sampling  assumption (and without the no-enhancement assumption);

b) taking into account undetected photons;

c) background events are taken into account;

d)  the {\it linear form of presentation} (non-negativity of a linear combination 
of coincidence and single rates).

\medskip

The latter feature (which is typically not emphasized in the literature) 
is crucial to find a simple procedure of optimization of experimental parameters 
and, hence, it makes the E-inequality the most promising experimental 
test to close the detection loophole and to reject local realism without 
the fair sampling assumption. Eberhard's optimization 
has two main outputs which play an important 
role in the experimental design:

\medskip

E1). It is possible to perform an experiment without fair sampling assumption for detection efficiency less than 82,8\%.
Nevertheless, detection efficiency must still be very high, at least 66.6\% (in the absence of background).

E2). The optimal parameters correspond to non-maximally entangled states. 
  
\medskip
   
In 2013, the possibility to proceed with overall efficiencies lower than 82.8\%  (but larger than 66.6\%) 
was explored for the E-inequality and the first  experimental test (``the Vienna test'') closing the detection  
loophole was published \cite{Zeilinger},  for more detailed presentation of statistical data see 
also \cite{Zeilinger2}, \cite{Zeilinger1}.

\subsection{Eberhard inequality and quantum mechanical probabilities}

Since the use of the Eberhard inequality is not common in quantum foundational studies, we present here in details calculation of 
quantum mechanical probabilities which violate it (for specially selected parameters of the experimental test). Here we follow 
the original paper of Eberhard, but we try to adapt the presentation for our purpose of improvement of optimization of parameters. 
Consider two detectors with the same efficiency $\eta$, which perform measurements in $N$ experiments. 
That is, in every experiment each of the detectors detects a photon in one of the trajectories with probability $\eta$. 
Let us construct density operators for particles in the ordinary and extraordinary beams. We will use the helicity basis
 for derivation of this operators.

Two main circular polarization states are described with the following vectors:
$
u = \frac{1}{\sqrt{2}}\begin{pmatrix}
1\\
-i
\end{pmatrix},\ 
v = \frac{1}{\sqrt{2}}\begin{pmatrix}
1\\
i
\end{pmatrix}.
$
They form the transformation matrix from standard basis to helicity basis:
$
W = \frac{1}{\sqrt{2}}\begin{pmatrix}
1 & 1\\
-i & i
\end{pmatrix}.
$
The inverse transformation can be made with the inverse matrix:
$
T = W^{-1} = \dfrac{1}{\sqrt{2}}\begin{pmatrix}
1 & i\\
1 & -1
\end{pmatrix}.
$
Then, consider a polarization prism which is rotated by an angle $\theta$. A particle which appeared in the ordinary beam has the following state:
$\psi_o = \begin{pmatrix}
\cos\theta\\
\sin\theta
\end{pmatrix}
$
and a particle in the extraordinary beam appeared with the state:
$
\psi_e = \begin{pmatrix}
-\sin\theta\\
\cos\theta
\end{pmatrix}.
$
In the helicity basis this states are described by the following equations, respectively:
$
\psi_o' = \dfrac{1}{\sqrt{2}}\begin{pmatrix}
e^{i\theta}\\
e^{-i\theta}
\end{pmatrix},\ 
\psi_e' = \dfrac{1}{\sqrt{2}}\begin{pmatrix}
ie^{i\theta}\\
-ie^{-i\theta}
\end{pmatrix}.
$
Then, corresponding density operators have the following form:
$
P_o = \psi_o' \cdot \psi_o'^{\dagger} = \dfrac{1}{2}\begin{pmatrix}
1 & e^{2i\theta}\\
e^{-2i\theta} & 1
\end{pmatrix},
$
$
P_e = \psi_e' \cdot \psi_e'^{\dagger} = \dfrac{1}{2}\begin{pmatrix}
1 & -e^{2i\theta}\\
-e^{-2i\theta} & 1
\end{pmatrix}.
$
Density operators of the considering system with two particles are described by tensor products. 
If the first particle appeared in the ordinary beam, it is described by $P_o \otimes I$, if the second particle appeared in the ordinary beam, 
then the corresponding operator is given by $I \otimes P_o$. Operators for particles in extraordinary beam can be constructed in the similar way.
Finally, putting it all together and using the formula of quantum expectation value, for the initial state of the system $\psi$ quantum 
mechanics predicts the following results:
\begin{eqnarray} \label{eq:Eberhard_n_start}
n_{oo}(\alpha_1, \beta_1) = N\frac{\eta^2}{4}\psi^\dagger[I + \sigma(\alpha_1)][I + \tau(\beta_1)]\psi,\\
n_{oe}(\alpha_1, \beta_2) = N\frac{\eta^2}{4}\psi^\dagger[I + \sigma(\alpha_1)][I - \tau(\beta_2)]\psi,\\
n_{ou}(\alpha_1, \beta_2) = N[\eta(1 - \eta)/2]\psi^\dagger[I + \sigma(\alpha_1)]\psi,\\ \label{eq:Eberhard_n_ou}
n_{eo}(\alpha_2, \beta_1) = N\frac{\eta^2}{4}\psi^\dagger[I - \sigma(\alpha_2)][I + \tau(\beta_1)]\psi,\\
n_{uo}(\alpha_2, \beta_1) = N[\eta(1 - \eta)/2]\psi^\dagger[I + \tau(\beta_1)]\psi,\\ \label{eq:Eberhard_n_uo}
n_{oo}(\alpha_2, \beta_2) = N\frac{\eta^2}{4}\psi^\dagger[I + \sigma(\alpha_2)][I + \tau(\beta_2)]\psi, \label{eq:Eberhard_n_finish}
\end{eqnarray}
where 
\[
\sigma(\alpha) = 
\begin{vmatrix}
0 & e^{2i(\alpha - \alpha_1)} & 0 & 0\\
e^{-2i(\alpha - \alpha_1)} & 0 & 0 & 0\\
0 & 0 & 0 & e^{2i(\alpha - \alpha_1)}\\
0 & 0 & e^{-2i(\alpha - \alpha_1)} & 0
\end{vmatrix}
\] 
and
\[
\tau(\beta) = 
\begin{vmatrix}
0 & 0 & e^{2i(\beta - \beta_1)} & 0\\
0 & 0 & 0 & e^{2i(\beta - \beta_1)}\\
e^{-2i(\beta - \beta_1)} & 0 & 0 & 0\\
0 & e^{-2i(\beta - \beta_1)} & 0 & 0
\end{vmatrix}.
\]
Note that matrices $I + \sigma(\alpha)$, $I - \sigma(\alpha)$, $I + \tau(\beta)$ and $I + \tau(\beta)$ can be simply derived from previously considered density operators with the properly choosen value of $\theta$. Also note that probability to fail to detect particle equals to $1 - \eta$ no matter what particle is considered, first of second.

Thus, we obtain the Eberhard inequality for quantum mechanical quantities:
\begin{multline*}
J_{\mathcal{B}}^{\mbox{ideal}} = n_{uo}(\alpha_2, \beta_1) + n_{eo}(\alpha_2, \beta_1) + n_{ou}(\alpha_1, \beta_2) \\
+ n_{oe}(\alpha_1, \beta_2) + n_{oo}(\alpha_2, \beta_2) - n_{oo}(\alpha_1, \beta_1) \geq 0.
\end{multline*}
However, in reality, in addition to correct detections during experiments some false positives may arise that is called 
background. In the Eberhard model it is assumed that the number of false positive detections for events of type $(o, o)$ can be ignored. 
We assume that the level of background does not depend on $\alpha$ and $\beta$, so for events $n_{uo}(\alpha_2, \beta_1) + 
n_{eo}(\alpha_2, \beta_1)$ and $n_{ou}(\alpha_1, \beta_2) + n_{oe}(\alpha_1, \beta_2)$, it 
has the same value $N\zeta$. The resulting inequality takes the form:
$
J_{\mathcal{B}} = J_{\mathcal{B}}^{\mbox{ideal}} + 2N\zeta \geq 0.
$

This inequality can be written as $\psi^\dagger\mathcal{B}\psi \geq 0$, where $\mathcal{B}$ is a matrix:
\[
\mathcal{B} = N \dfrac{\eta}{2}
\begin{vmatrix}
2 - \eta + \xi & 1 - \eta & 1 - \eta & A^*B^* - \eta\\
1 - \eta & 2 - \eta + \xi & AB^* - \eta & 1 - \eta\\
1 - \eta & A^*B - \eta & 2 - \eta + \xi & 1 - \eta\\
AB - \eta & 1 - \eta & 1 - \eta & 2 - \eta + \xi
\end{vmatrix},
\]
where $A = \eta/2(e^{2i(\alpha_1-\alpha_2)} - 1)$, $B = e^{2i(\beta_1 - \beta_2)} -1$, $\xi = 4\zeta/\eta$.

For the implementation of the experiment, which could show a violation of this inequality, we will search 
for the parameters such that $J_{\mathcal{B}} < 0$. Consider the case when $\zeta = 0$, that is, the detectors 
do not give false positives, and $\alpha_1 - \alpha_2 = \beta_1 - \beta_2 = \theta$. We will use the following quantum state:
$
\psi = \frac{1}{2\sqrt{1 + r^2}}
\begin{vmatrix}
(1+r)e^{-i\omega}\\
-(1 - r)\\
-(1 - r)\\
(1 + r)e^{i\omega}
\end{vmatrix},
$
where $0 \leq r \leq 1$, $\alpha_1 = \omega / 2 - 90^\circ$ and $\beta_1 = \omega / 2$.

\section{The results of optimization of parameters for 
experimental tests based on Eberhard's inequality}

Our numerical optimization of parameters of the experimental tests to violate Eberhard's 
inequality will be based on {\it the Nelder-Mead optimization method.} 
The Nelder–Mead method  \cite{NM} (also known as downhill simplex method)
is widely used for nonlinear optimization problems. This  numerical method is typically applied  
to problems for which derivatives may not be known. Its applications are especially successful in the case 
of multi-dimensional spaces of parameters. 

In this section we present a part of the results of our studies, the results of numerical optimization of parameters to violate Eberhard's 
inequality as much as possible. We start with comparison with the 
original Eberhard model \cite{Eberhard}, then we consider the case of detectors having different efficiencies,
so in general $\eta_1\not= \eta_2.$ Finally, we consider the model, see Kofler et al. \cite{Zeilinger1} which was
used in the recent Bell test \cite{Zeilinger} based on the Eberhard inequality, the ``Vienna-13 experiment''   

\subsection{Optimization of parameters for the Eberhard model}
\label{EBM}

For every $\eta$ let us find parameters $r, \omega, \theta$ that allows the inequality to be violated most strongly.
To do so, we will minimize the  function f=$J_{\mathcal{B}}(r, \omega, \theta) / N$ using the Nelder-Mead method.  
\begin{table}[h]
\centering
\begin{tabular}{|c|c|c|c|c|}
\hline
$\eta$ & $r$ & $\omega,^\circ$ & $\theta,^\circ$ & $J_{\mathcal{B}} / N$
\\\hline
0.7 & 0.136389 & 3.40081 & 21.4266 & -0.000453562\\\hline
0.75 & 0.310518 & 9.73143 & 31.9603 & -0.00615095\\\hline
0.8 & 0.465228 & 14.8979 & 37.9215 & -0.02191\\\hline
0.85 & 0.607424 & 18.5808 & 41.5341 & -0.0496902\\\hline
0.9 & 0.741202 & 20.9153 & 43.6381 & -0.0899078\\\hline
0.95 & 0.87067 & 22.141 & 44.6958 & -0.142436\\\hline
1 & 0.999997 & 22.5 & 45 & -0.207107\\\hline
\end{tabular}
\caption{Optmimal parameters values for $J_{\mathcal{B}} / N$ from the Eberhard inequality}
\label{tab:Eberhard_repeat}
\end{table}

The values obtained while optimizing $J_{\mathcal{B}}(r, \omega, \theta) / N$ are shown in  
Table~\ref{tab:Eberhard_repeat}. During the optimization process  the $\zeta$-values were set to zero, 
because this parameter brings a constant contribution $2\zeta$ into the $J_{\mathcal{B}} / N$ value. It increases the 
$J_{\mathcal{B}} / N$ value by a constant regardless of other parameters, 
therefore not affecting optimization result. Thus values from Table~\ref{tab:Eberhard_repeat} 
match with values obtained by Eberhard \cite{Eberhard} for non-zero context level, and parameters such that the inequality is
violated the most strongly for $\zeta = 0$ match with parameters such that the inequality is
violated and $\zeta$ has the maximum value.

\begin{figure}[h]
\includegraphics[scale=0.3]{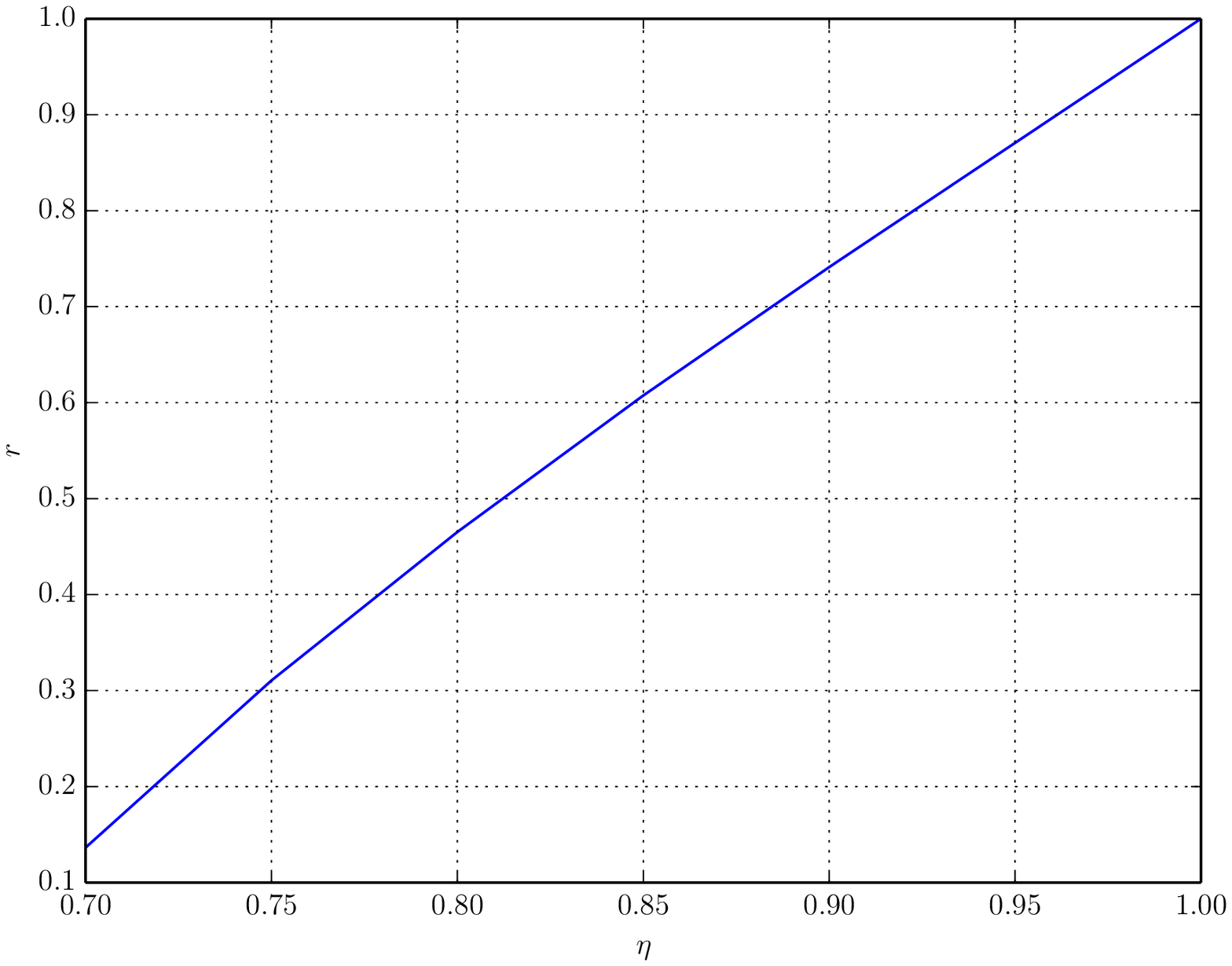}
\includegraphics[scale=0.3]{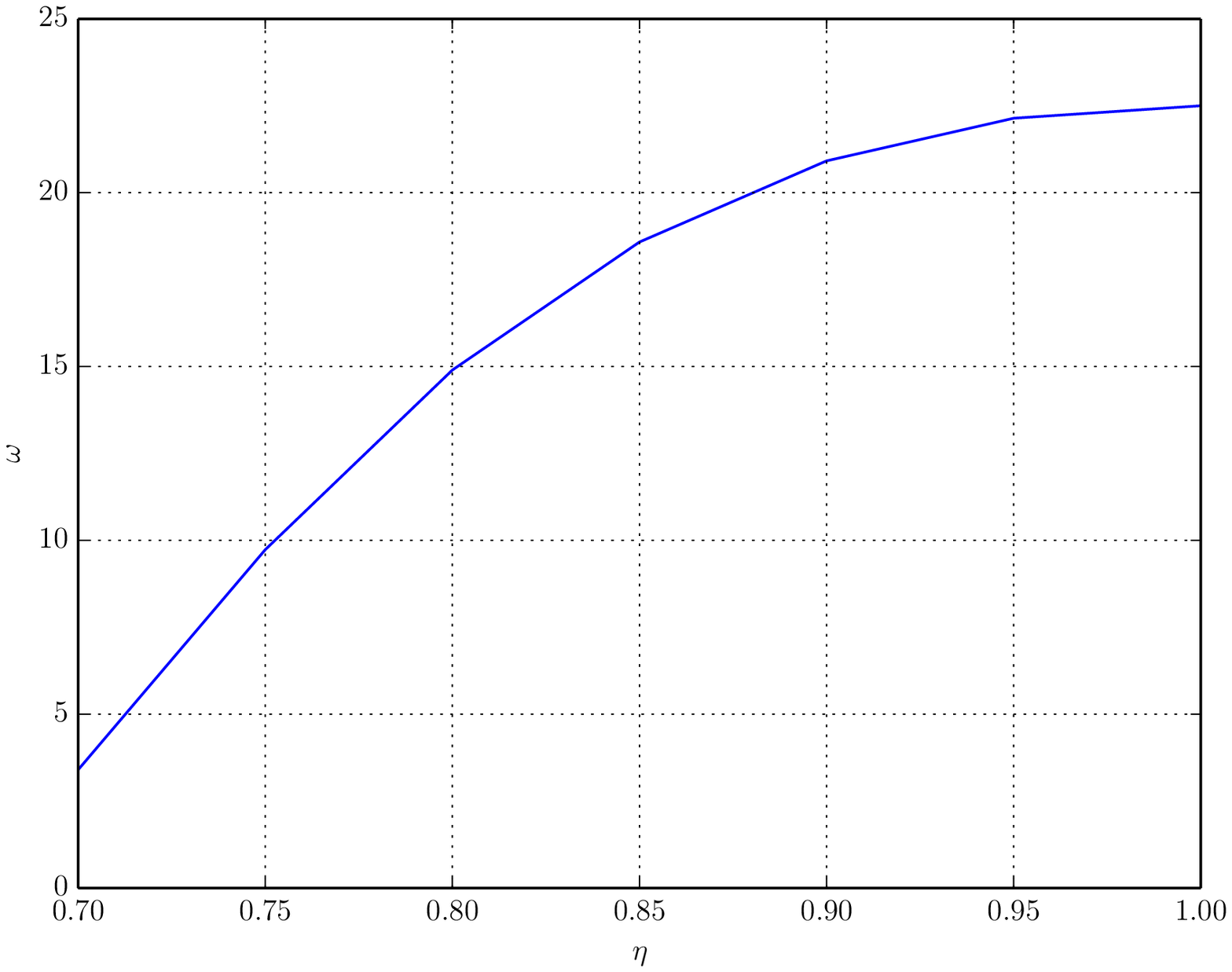}
\caption{Optimized $r$ and $\omega$ values for different detectors efficiency values}
\label{fig:psi_opt}
\end{figure}

\begin{figure}[h]
\includegraphics[scale=0.3]{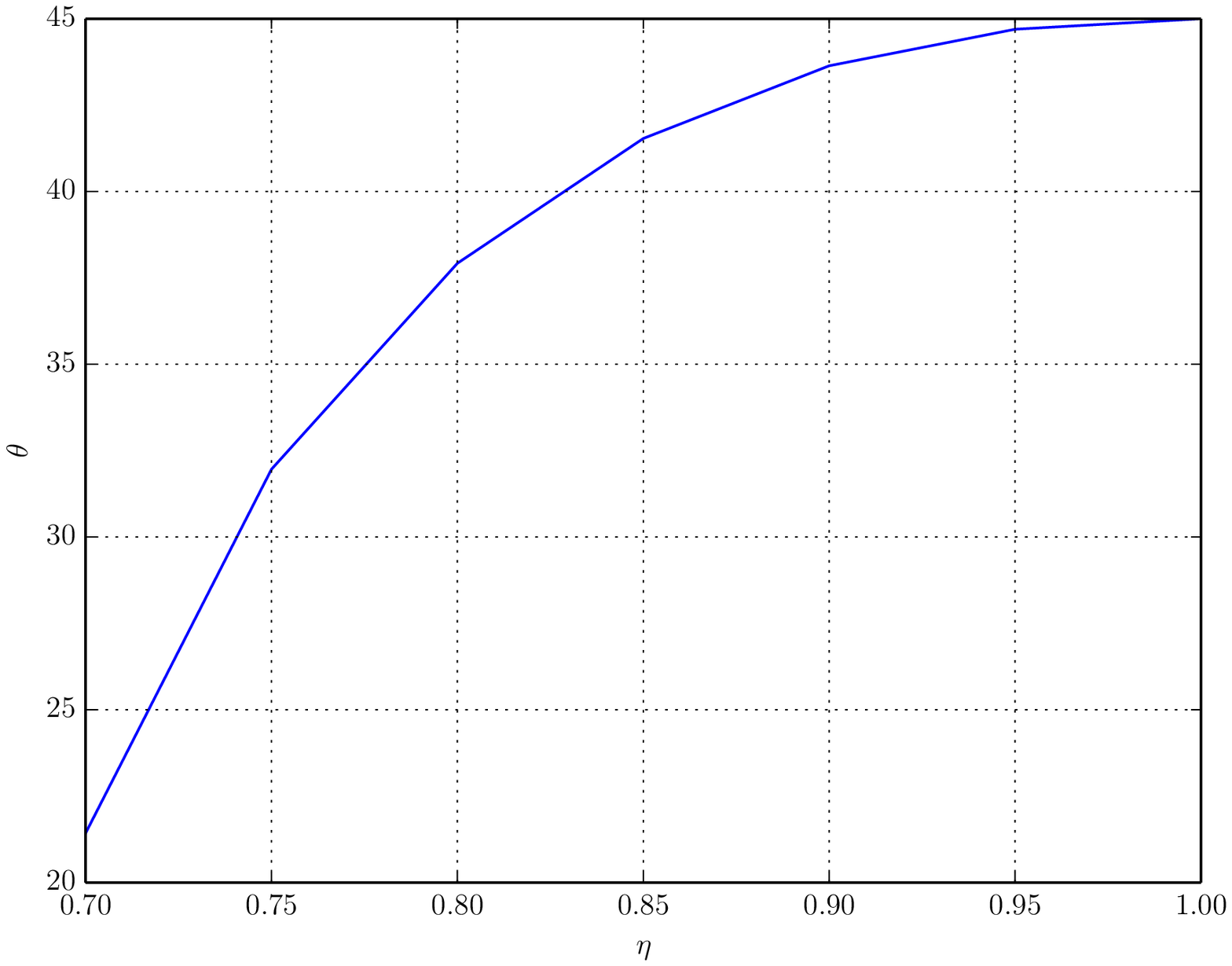}
\includegraphics[scale=0.3]{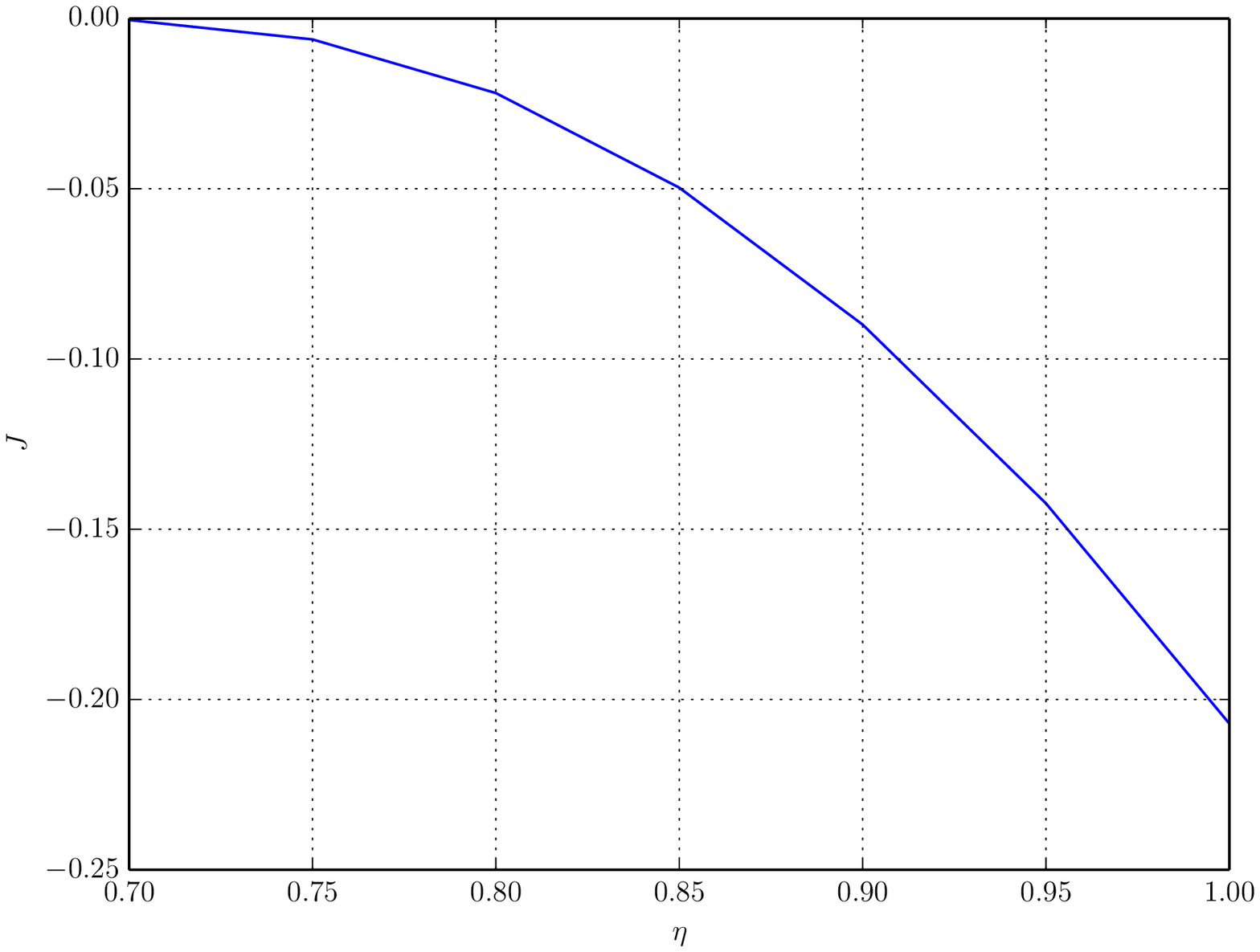}
\caption{Optimized $\theta$ qnd $J_{\mathcal{B}}/N$ values for different detector efficiency values}
\label{fig:J_opt}
\end{figure}

At Fig.  \ref{fig:psi_opt} relations between obtained parameters $r$ and $\omega$ and the efficiency $\eta$  
are shown, at the first picture of Fig. \ref{fig:J_opt} 
relations between obtained parameter $\theta$ and efficiency 
is shown, at the second picture of this figure dependence of the minimum function values  on efficiency is 
presented.

\subsection{Optimization of parameters for detectors with different efficiencies; Eberhard model}

In real experiments detectors more often have different efficiency values, the
formulas \eqref{eq:Eberhard_n_start}-\eqref{eq:Eberhard_n_finish} can be easily adapted to this case.
The Eberhard inequality can be written as $\psi^\dagger\mathcal{B}\psi \geq 0$, where $\mathcal{B}$ is a matrix:
\[
\mathcal{B} = \dfrac{N}{2}
\begin{vmatrix}
C + \xi & \eta_1(1 - \eta_2) & \eta_2(1 - \eta_1) & A^*B^* - \eta_1\eta_2\\
\eta_1(1 - \eta_2) & C + \xi & AB^* - \eta_1\eta_2 & \eta_2(1 - \eta_1)\\
\eta_2(1 - \eta_1) & A^*B - \eta_1\eta_2 & C + \xi & \eta_1(1 - \eta_2)\\
AB - \eta_1\eta_2 & \eta_2(1 - \eta_1) & \eta_1(1 - \eta_2) & C + \xi
\end{vmatrix},
\]
where $A = \eta_1/2(e^{2i(\alpha_1-\alpha_2)} - 1)$, $B = \eta_2(e^{2i(\beta_1 - \beta_2)} -1)$, $C = \eta_1 + \eta_2 - \eta_1\eta_2$ and $\xi = 4\zeta$.

\begin{figure}[h]
\includegraphics[scale=0.3]{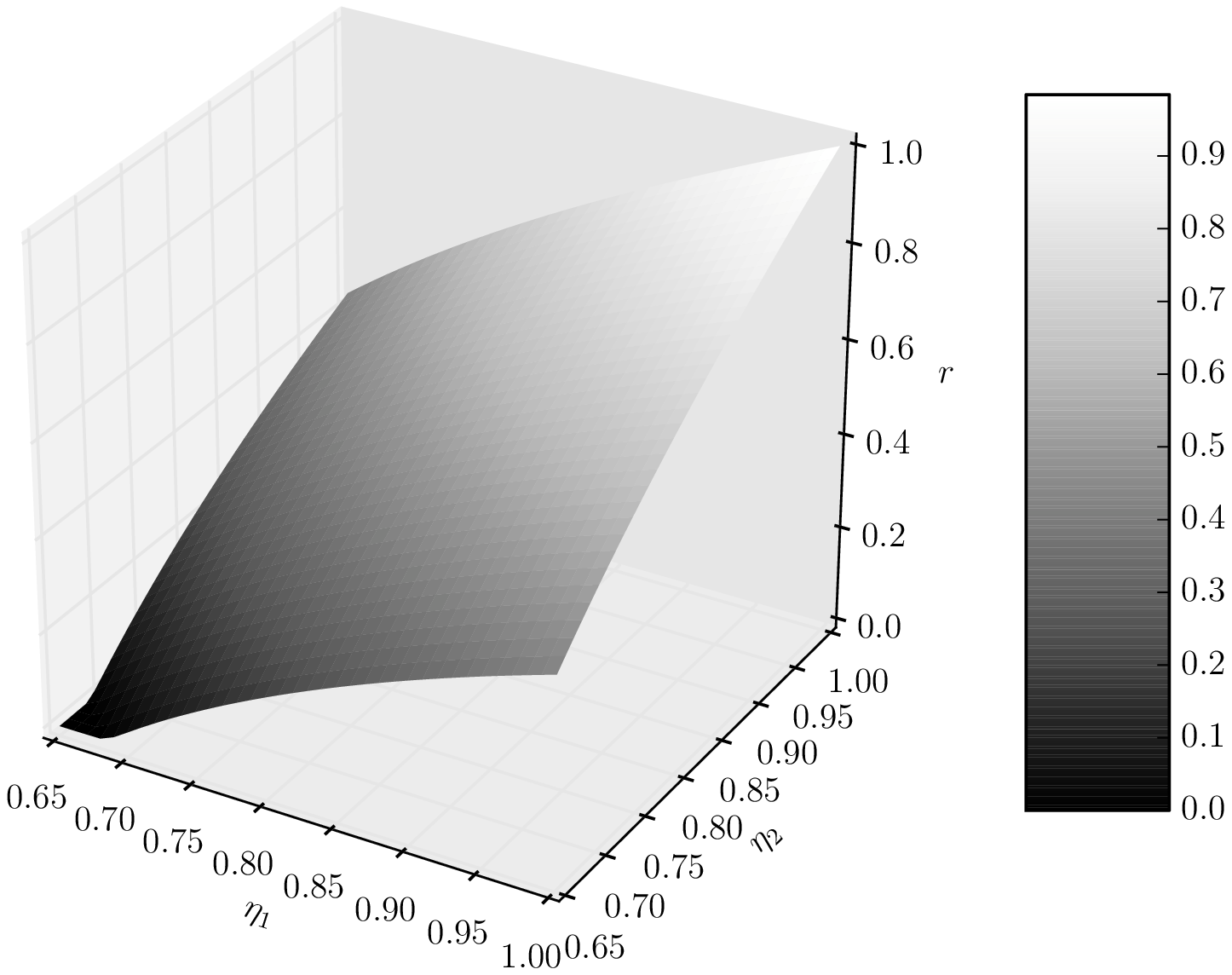}
\includegraphics[scale=0.3]{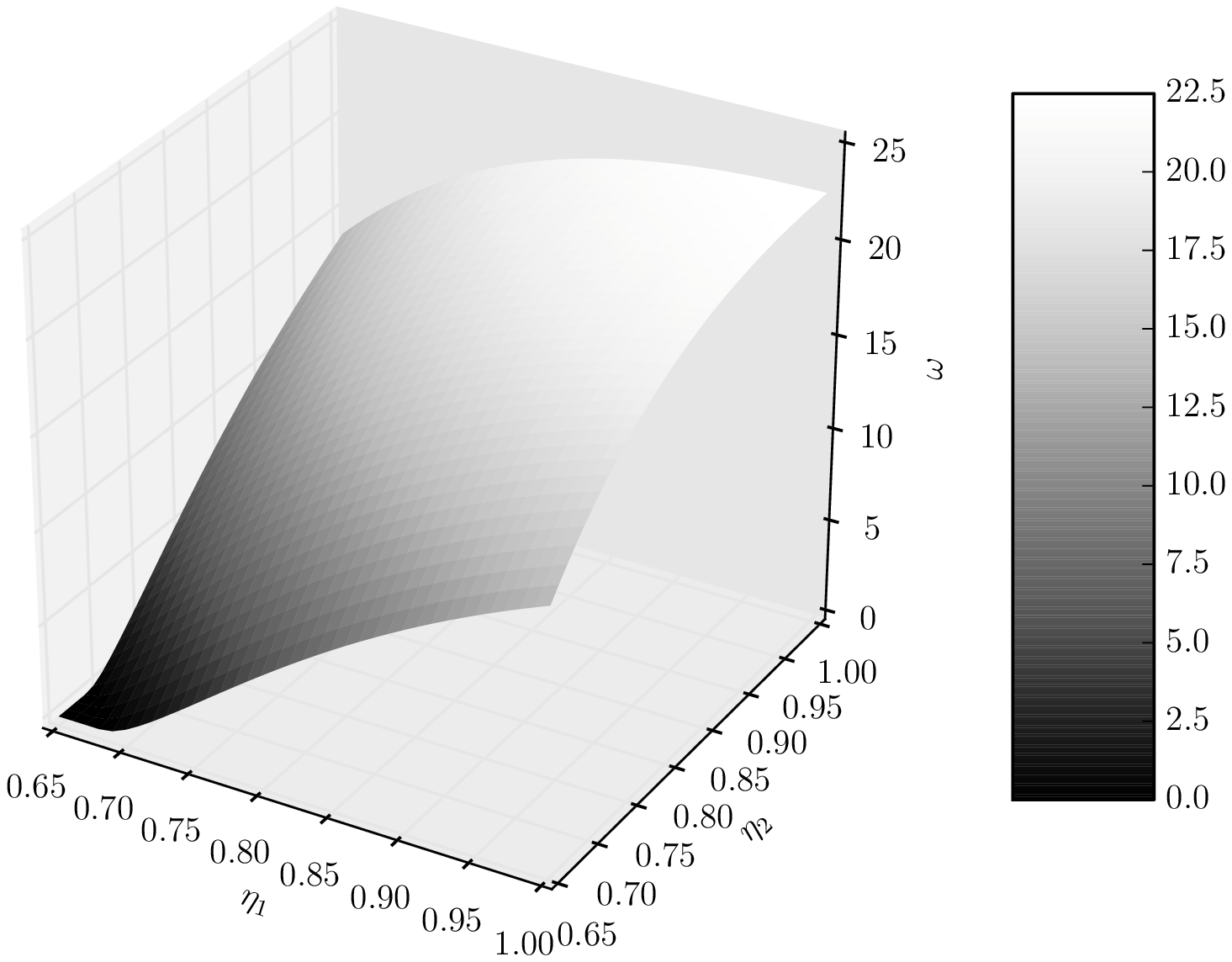}
\caption{Optimized $r$ and $\omega$ values for different detector efficiencies}
\label{fig:psi_opt_3d}
\end{figure}

\begin{figure}[t]
\includegraphics[scale=0.3]{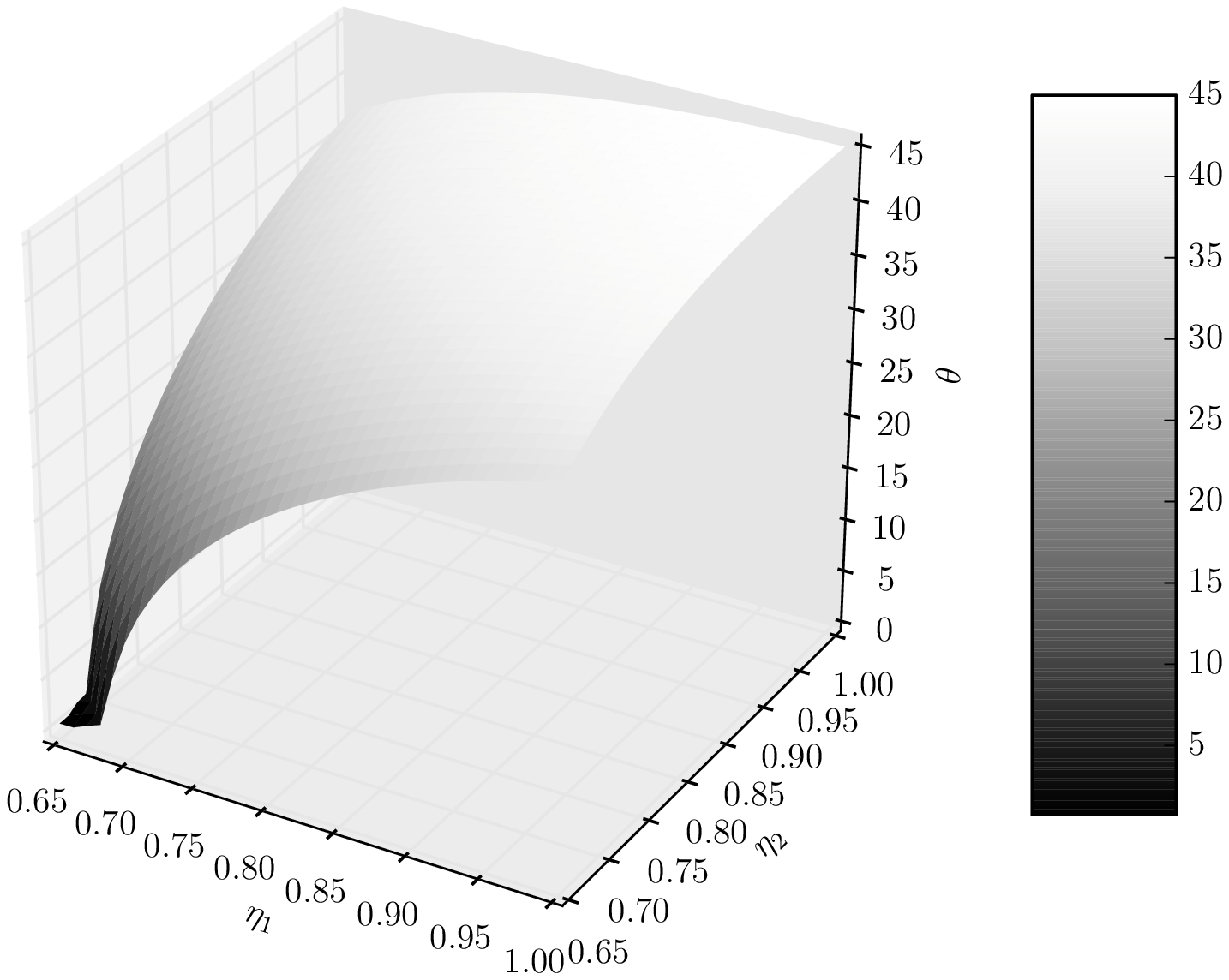}
\includegraphics[scale=0.3]{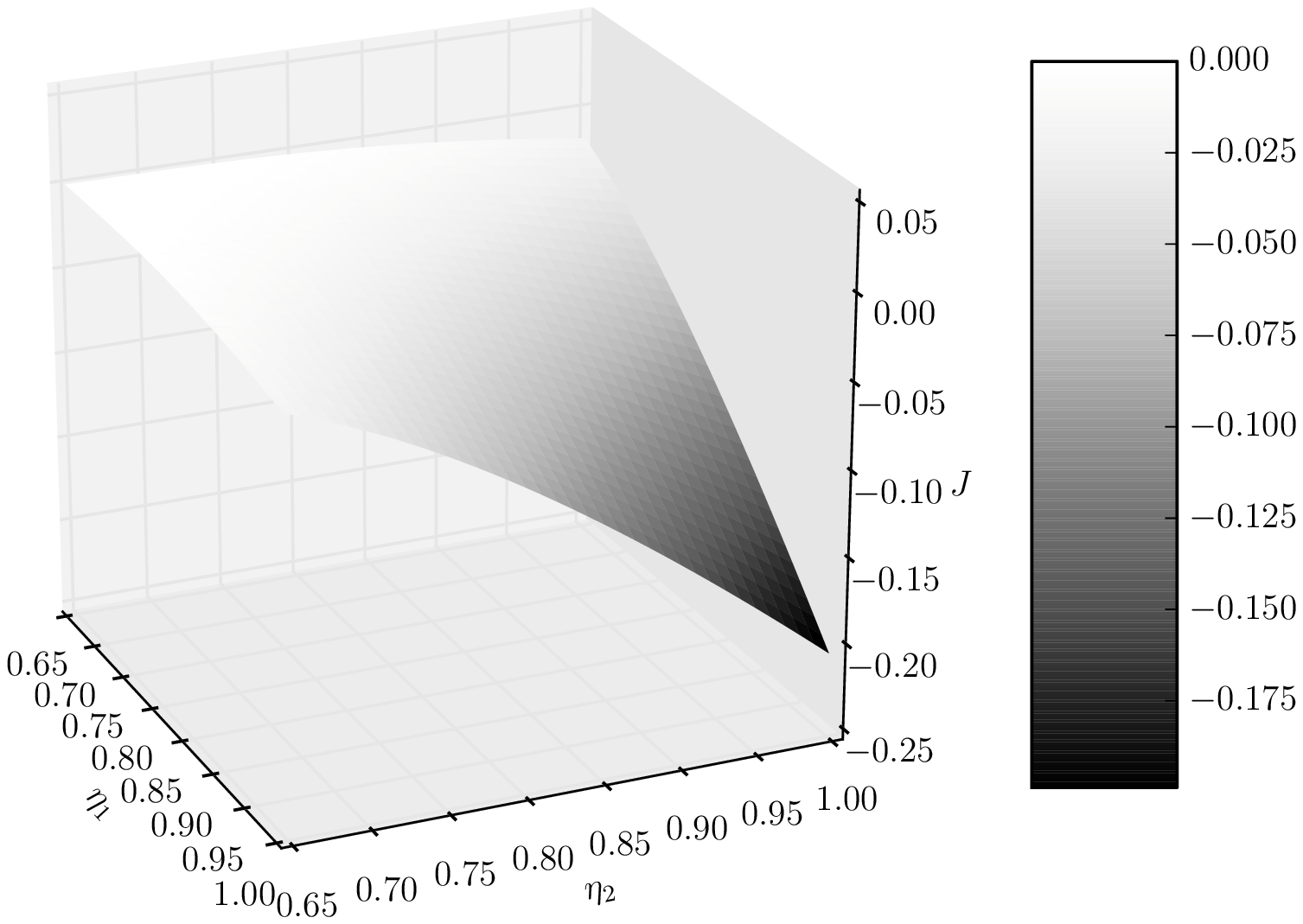}
\caption{Optimized values of $\theta$ and $J_{\mathcal{B}}/N$ for different detector efficiencies}
\label{fig:J_opt_3d}
\end{figure}

At Fig.\ref{fig:psi_opt_3d} - \ref{fig:J_opt_3d} optimal parameters are shown, together with 
the minimized function for different detector efficiencies.

\begingroup
\small

\begin{longtable}{|c|c|c|c|c|c|c|}
\hline
$\eta_1$ & $\eta_2$ & $r$ & $\omega,^\circ$ & $\theta,^\circ$ & $J_{\mathcal{B}}/N$ & $\zeta$
\\\hline
\endfirsthead

\hline
$\eta_1$ & $\eta_2$ & $r$ & $\omega,^\circ$ & $\theta,^\circ$ & $J_{\mathcal{B}}/N$ & $\zeta$
\\\hline
\endhead

\endfoot

\endlastfoot

\hline
0.65 & 0.65 & 9.73816e-05 & 5.32909e-05 & 0.609422 & 4.13366e-10 & --\\\hline
0.65 & 0.7 & 0.0325726 & 0.4367 & 10.3918 & -5.37452e-06 & 2.68726e-06\\\hline
0.65 & 0.75 & 0.122804 & 2.94407 & 20.3249 & -0.000327839 & 0.00016392\\\hline
0.65 & 0.8 & 0.199412 & 5.64399 & 25.8864 & -0.00152446 & 0.000762231\\\hline
0.65 & 0.85 & 0.266448 & 8.1196 & 29.7694 & -0.00385363 & 0.00192682\\\hline
0.65 & 0.9 & 0.326193 & 10.2951 & 32.6807 & -0.00737942 & 0.00368971\\\hline
0.65 & 0.95 & 0.380046 & 12.1725 & 34.9409 & -0.0120653 & 0.00603265\\\hline
0.65 & 1 & 0.42905 & 13.7822 & 36.7384 & -0.0178271 & 0.00891353\\\hline
0.7 & 0.65 & 0.0325721 & 0.436691 & 10.3917 & -5.37452e-06 & 2.68726e-06\\\hline
0.7 & 0.7 & 0.136389 & 3.40081 & 21.4266 & -0.000453562 & 0.000226781\\\hline
0.7 & 0.75 & 0.223629 & 6.53572 & 27.3754 & -0.00217282 & 0.00108641\\\hline
0.7 & 0.8 & 0.299639 & 9.33741 & 31.4432 & -0.00551773 & 0.00275887\\\hline
0.7 & 0.85 & 0.367155 & 11.7325 & 34.4284 & -0.0105412 & 0.00527059\\\hline
0.7 & 0.9 & 0.427895 & 13.7454 & 36.6985 & -0.0171516 & 0.0085758\\\hline
0.7 & 0.95 & 0.48304 & 15.4224 & 38.4617 & -0.0251977 & 0.0125989\\\hline
0.7 & 1 & 0.533433 & 16.8118 & 39.8493 & -0.034513 & 0.0172565\\\hline
0.75 & 0.65 & 0.122804 & 2.94407 & 20.3249 & -0.000327839 & 0.00016392\\\hline
0.75 & 0.7 & 0.223629 & 6.53572 & 27.3754 & -0.00217282 & 0.00108641\\\hline
0.75 & 0.75 & 0.310518 & 9.73143 & 31.9603 & -0.00615095 & 0.00307547\\\hline
0.75 & 0.8 & 0.387235 & 12.4151 & 35.2194 & -0.0123604 & 0.0061802\\\hline
0.75 & 0.85 & 0.455977 & 14.6188 & 37.6299 & -0.0206635 & 0.0103318\\\hline
0.75 & 0.9 & 0.518149 & 16.4057 & 39.45 & -0.0308332 & 0.0154166\\\hline
0.75 & 0.95 & 0.574864 & 17.8435 & 40.8423 & -0.0426257 & 0.0213128\\\hline
0.75 & 1 & 0.626962 & 18.9919 & 41.9138 & -0.0558135 & 0.0279068\\\hline
0.8 & 0.65 & 0.199412 & 5.64399 & 25.8864 & -0.00152446 & 0.000762231\\\hline
0.8 & 0.7 & 0.299639 & 9.33741 & 31.4432 & -0.00551773 & 0.00275887\\\hline
0.8 & 0.75 & 0.387235 & 12.4151 & 35.2194 & -0.0123604 & 0.0061802\\\hline
0.8 & 0.8 & 0.465228 & 14.8979 & 37.9215 & -0.02191 & 0.010955\\\hline
0.8 & 0.85 & 0.535462 & 16.8648 & 39.9011 & -0.0338613 & 0.0169307\\\hline
0.8 & 0.9 & 0.59933 & 18.4036 & 41.3692 & -0.0478775 & 0.0239387\\\hline
0.8 & 0.95 & 0.657889 & 19.5935 & 42.4623 & -0.063646 & 0.031823\\\hline
0.8 & 1 & 0.712001 & 20.5014 & 43.274 & -0.0808982 & 0.0404491\\\hline
0.85 & 0.65 & 0.266448 & 8.1196 & 29.7694 & -0.00385363 & 0.00192682\\\hline
0.85 & 0.7 & 0.367155 & 11.7325 & 34.4284 & -0.0105412 & 0.00527059\\\hline
0.85 & 0.75 & 0.455977 & 14.6188 & 37.6299 & -0.0206635 & 0.0103318\\\hline
0.85 & 0.8 & 0.535462 & 16.8648 & 39.9011 & -0.0338613 & 0.0169307\\\hline
0.85 & 0.85 & 0.607424 & 18.5808 & 41.5341 & -0.0496902 & 0.0248451\\\hline
0.85 & 0.9 & 0.673214 & 19.8694 & 42.7109 & -0.0677295 & 0.0338647\\\hline
0.85 & 0.95 & 0.733924 & 20.817 & 43.552 & -0.087619 & 0.0438095\\\hline
0.85 & 1 & 0.790464 & 21.4958 & 44.1427 & -0.109064 & 0.0545318\\\hline
0.9 & 0.65 & 0.326193 & 10.2951 & 32.6807 & -0.00737942 & 0.00368971\\\hline
0.9 & 0.7 & 0.427895 & 13.7454 & 36.6985 & -0.0171516 & 0.0085758\\\hline
0.9 & 0.75 & 0.518149 & 16.4057 & 39.45 & -0.0308332 & 0.0154166\\\hline
0.9 & 0.8 & 0.59933 & 18.4036 & 41.3692 & -0.0478775 & 0.0239387\\\hline
0.9 & 0.85 & 0.673214 & 19.8694 & 42.7109 & -0.0677295 & 0.0338647\\\hline
0.9 & 0.9 & 0.741202 & 20.9153 & 43.6381 & -0.0899078 & 0.0449539\\\hline
0.9 & 0.95 & 0.804477 & 21.6349 & 44.2627 & -0.11402 & 0.0570101\\\hline
0.9 & 1 & 0.863896 & 22.1002 & 44.661 & -0.139755 & 0.0698776\\\hline
0.95 & 0.65 & 0.380046 & 12.1725 & 34.9409 & -0.0120653 & 0.00603265\\\hline
0.95 & 0.7 & 0.48304 & 15.4224 & 38.4617 & -0.0251977 & 0.0125989\\\hline
0.95 & 0.75 & 0.574864 & 17.8435 & 40.8423 & -0.0426257 & 0.0213128\\\hline
0.95 & 0.8 & 0.657889 & 19.5935 & 42.4623 & -0.063646 & 0.031823\\\hline
0.95 & 0.85 & 0.733924 & 20.817 & 43.552 & -0.087619 & 0.0438095\\\hline
0.95 & 0.9 & 0.804477 & 21.6349 & 44.2627 & -0.11402 & 0.0570101\\\hline
0.95 & 0.95 & 0.87067 & 22.141 & 44.6958 & -0.142436 & 0.0712182\\\hline
0.95 & 1 & 0.933431 & 22.4101 & 44.924 & -0.172546 & 0.086273\\\hline
1 & 0.65 & 0.42905 & 13.7822 & 36.7384 & -0.0178271 & 0.00891353\\\hline
1 & 0.7 & 0.533433 & 16.8118 & 39.8493 & -0.034513 & 0.0172565\\\hline
1 & 0.75 & 0.626962 & 18.9919 & 41.9138 & -0.0558135 & 0.0279068\\\hline
1 & 0.8 & 0.712001 & 20.5014 & 43.274 & -0.0808982 & 0.0404491\\\hline
1 & 0.85 & 0.790464 & 21.4958 & 44.1427 & -0.109064 & 0.0545318\\\hline
1 & 0.9 & 0.863896 & 22.1002 & 44.661 & -0.139755 & 0.0698776\\\hline
1 & 0.95 & 0.933431 & 22.4101 & 44.924 & -0.172546 & 0.086273\\\hline
1 & 1 & 0.999997 & 22.5 & 45 & -0.207107 & 0.103553\\\hline
\caption{Optimal parameters values for the case with different detector efficiencies}
\label{tab:different_etas}
\end{longtable}

\endgroup

Parameters values obtained during optimization process along with maximal allowable noise level 
are shown in Table~\ref{tab:different_etas}. The bigger each value of efficiency separately, the more 
strongly inequality can be violated. Minimal efficiency values such that the violation is possible are close 
to  $\eta_1 = \eta_2 = 0.67$ value, matching Eberhard results.

Generally, for every state $\psi$ that minimize an expectation value the following corollary holds.

{\bf Theorem.}
{\it Quantum state $\psi$ minimizing target function $J$ is an eigenvector of the matrix $\mathcal{B}$ and the dispersion for it is equal to zero.}

This theorem can be proved using Courant-Fischer theorem.

{\bf Theorem}[Courant-Fisher]
{\it Let $A$ be a $n \times n$ Hermitian matrix with eigenvalues $\lambda_1 \leq \lambda_2 \leq \ldots \leq \lambda_n$. Rayleigh–Ritz quotient for this matrix is defined by 
\[
R_A(x) = \frac{(Ax, x)}{(x,x)}.
\]
For $1 \leq k \leq n$, let $S_k$ denote the span of $v_1,\ldots ,v_k$ and let $Sk$ denote the orthogonal complement of $S_k$.
Then
\[
\lambda_1 \leq R_A(x) \leq \lambda_n, \ \forall x \in \mathbb{C}^n\setminus \{0\}
\]
and}
\[
\lambda_k = \max\{\min \{R_A(x)\mid x\neq 0 \in U\} \mid \dim(U) = k \},
\]
\[
\lambda_k = \min \{ \max\{R_A(x)\mid x\neq 0 \in U\} \mid  \dim(U) = n - k +1\}
\]

It means that the obtained states are optimal not only from the mathematical expectation point of view, but also from 
possible spread of measurement results point of view expressed in terms of dispersion.

\subsection{Optimization of parameters in the model for the Vienna-13 experiment}

To match the real experimental situation, see Gustina et al. \cite{Zeilinger} (Vienna-13 experiment), 
in article of Kofler et al. \cite{Zeilinger1} analysis of the use of the Eberhard inequality in this concrete experiment 
was performed. This analysis led to the conclusion that data produced in the Vienna-13 experiment  \cite{Zeilinger} is described by a more complicated
model (in the standard quantum framework)\footnote{As one of the aims, the work of Kofler et al. \cite{Zeilinger1}  
has justification of the statistical output of  the Vienna-13 experiment by using standard quantum mechanical tools. 
This was questioned by the author of the paper \cite{Q}.}
than the original Eberhard model \cite{Eberhard}. This does not decrease the value of the original Eberhard
study. Kofler et al. \cite{Zeilinger1} just pointed that some important additional ``technicalities'' have to be taken into account.

During experiments \cite{Zeilinger} the values of the quantities $n_{ou}, n_{uo}$  were found by the following formulas:
\[
n_{ou}(\alpha_1, \beta_2) = S_o^A(\alpha_1) - n_{oo}(\alpha_1, \beta_2) - n_{oe}(\alpha_1, \beta_2),
\]
\[
n_{uo}(\alpha_2, \beta_1) = S_o^B(\beta_1) - n_{oo}(\alpha_2, \beta_1) - n_{eo}(\alpha_2, \beta_1),
\]
where $S_o^A, S_o^B$ -- an amount of clicks in the ordinary beam for the first and second system 
correspondingly.

In this case the Eberhard inequality takes form:
\begin{equation}
J = -n_{oo}(\alpha_1, \beta_1) + S_o^A(\alpha_1) - n_{oo}(\alpha_1, \beta_2) + S_o^B(\beta_1) - n_{oo}(\alpha_2, \beta_1) + n_{oo}(\alpha_2, \beta_2) \geq 0
\label{eq:Zeilinger_J}
\end{equation}

To model the output of the Vienna-13 experiment, one cannot proceed, as Eberhard did, with pure states.  
Consider a density operator as a quantum state of the system:
\[
\rho = \frac{1}{\sqrt{1+r^2}}
\begin{vmatrix}
0 & 0 & 0 & 0\\
0 & 1 & Vr & 0\\
0 & Vr & r^2 & 0\\
0 & 0 & 0 & 0
\end{vmatrix},
\]
where $0 \leq r \leq 1$ and $0 \leq V \leq 1$.
In this case predictions of quantum mechanics for values included into the inequality become:
\begin{eqnarray*}
\tilde{S}_o^A(\alpha_i) = \eta_1 N \Tr[\rho(\hat{P}_A(\alpha_i) \otimes I)]\\
\tilde{S}_o^B(\beta_i) = \eta_1 N \Tr[\rho(I \otimes \hat{P}_B(\beta_i))]\\
\tilde{n}_{oo}(\alpha_i, \beta_i) = \eta_1\eta_2 N \Tr[\rho(\hat{P}_A(\alpha_i) \otimes \hat{P}_B(\beta_i))],
\end{eqnarray*}
where $\hat{P}_A, \hat{P}_B$ -- projection operators on ordinary beam direction for the first and second prisms:
\[
\hat{P}(\gamma) = 
\begin{pmatrix}
\cos^2\gamma & \cos\gamma\sin\gamma\\
\cos\gamma\sin\gamma & \sin^2\gamma
\end{pmatrix}.
\]

With regard to false clicks during time $T$ $S_o^A, S_o^B$ values take the following form:
\[
S_o^A(\alpha_i) = \tilde{S}_o^A(\alpha_i) + \zeta T,
\]
\[
S_o^B(\beta_i) = \tilde{S}_o^B(\beta_i) + \zeta T.
\]

Besides noise,  the model of the Vienna-13 experiment, see Kofler et al. \cite{Zeilinger1}, also considers inconsistencies in time when pairs from different 
launches are detected as a conjugate events. Let us introduce temporary window value $\tau_c$, within 
which conjugate events must be detected. In this case $n_{oo}(\alpha_i, \beta_i)$ can be found using 
the following formulas:
\[
n_{oo}(\alpha_i, \beta_i) = \tilde{n}_{oo}(\alpha_i, \beta_i) + n_{oo}^{acc}(\alpha_i, \beta_i),
\]
\[
n_{oo}^{acc}(\alpha_i, \beta_i) = S_o^A(\alpha_i)S_o^B(\beta_i)\frac{\tau_c}{T}\left(1 - \frac{\tilde{n}_{oo}(\alpha_i, \beta_i)}{S_o^A(\alpha_i)}\right) \left( 1 - \frac{\tilde{n}_{oo}(\alpha_i, \beta_i)}{S_o^B(\beta_i)}\right).
\]

For this model, optimization for the quantity $J$ given by the expression \eqref{eq:Zeilinger_J} was performed, and,
for the selected values of experimental parameters 
($\eta_1, \eta_2, r, V, T, \tau_c, N, \zeta$), the values of the angles that
minimize the target function were found. In particular, we remark that Gustina et al. \cite{Zeilinger} approached the following 
levels of  detectors efficiencies: $\eta_1= 73.77$ and $\eta_2= 78.59.$

\begin{table}
\begin{tabular}{|c|c|c|c|c|c|}
\hline 
 & $\alpha_1, ^\circ$ & $\alpha_2, ^\circ$ & $\beta_1, ^\circ$ & $\beta_2, ^\circ$ & $J$ \\ 
\hline 
Results from the article & 85.6 & 118.0 & -5.4 & 25.9 & -120191 \\ 
\hline 
Results of the optimization & 85.0 & 115.1 & -4.0 & 27.4 & -126060 \\ 
\hline 
\end{tabular}
\caption{Optimal parameters comparison with ones from Zeilinger article \cite{Zeilinger}}
\label{tab:Zeilinger_parameters}
\end{table}

Optimization results are shown in Table~\ref{tab:Zeilinger_parameters}. According to obtained values, 
optimal angle values for prism installation differ from given in \cite{Zeilinger} and let inequality to be violated 
stronger.  We also point out that asymmetry in detectors' contributions leads to a possibility to play with this asymmetry.
In particular, we found that if experimentalists who did the Vienna-13 experiment were simply permuted the detectors,
they would get a stronger violation: $J= -123050.$

\section{Optimization of parameters for randomly fluctuating angles of polarization beam splitters}
\label{last}

In the Eberhard model \cite{Eberhard} and  the model for the Vienna-13 experiment \cite{Zeilinger}   optimization of 
experimental parameters was performed 
under the assumption that the angles of polarization beam splitters can be chosen exactly. The optimization gives some 
concrete values and it was assumed that experimentalists can setup the experimental design with precisely these angles. 
However, this assumption does not match the real experimental situation. Although the precision of selection of angles
of polarization beam splitters is very high (e.g., in the Vienna-13 experiment  \cite{Zeilinger} -- private communication), 
nevertheless, there are errors which 
can lead to deviations from the expected value of the   $f = J_{\mathcal{B}}(r, \omega, \theta).$ Therefore it is important 
to study the problem of the statistical stability of optimization with respect to random fluctuations of the angles. In this 
section we present the corresponding theoretical considerations, the results of numerical optimization (again with the aid of 
Nelder-Mead method) will be presented in section \ref{last}. 

Taking into account possible random fluctuations of the angles makes the question of optimization more complicated. As the result 
of such fluctuations, in the optimal point for the mathematical expectation the dispersion is nontrivial. In principle, one can get 
a large magnitude of the absolute value of the mathematical expectation, but at the same time also  a large magnitude of the standard
deviation. Therefore it is natural to optimize not simply the mathematical expectation 
given by the function $J_{\mathcal{B}}$, but the quantity $J_{\mathcal{B}}/\sigma.$      

In signal processing the quantity $$K=\mu/\sigma,$$ where $\mu$ is average and $\sigma$ is the standard deviation, is widely 
used and known as {\it signal/noise ratio} (SNR), see \cite{CI1}, \cite{CI2}:
 This interpretation can be used even in our framework (if we interpret random fluctuations
of angles as generated by a kind of noise), although we operate not with 
continuous signals, but with the discrete clicks of detectors.  
We also remark that SNR is as the reciprocal of {\it the coefficient of variation,}
$\sigma/\mu.$ It shows the extent of variability in relation to mean of the sample.

One of specialties of our work with SNR or the coefficient of variation is that in the 
standard situations they are used only for measurements with nonnegative values.
In our case values are negative. However, we can simply change the sign of measurement 
quantity. Therefore we proceed with negative K by taking into account that statistical meaning 
has to be assigned to its absolute value -- the reciprocal of the 
the relative standard deviation (RSD) which is the absolute value of the coefficient of variation,
$\vert \sigma/\mu \vert.$ 

We now move to theoretical modeling of randomly fluctuating angles of polarization beam splitters.
Generally any self-adjoint quantum operator $A$ can be represented using spectral decomposition as 
$A = \int_{-\infty}^{+\infty}\lambda dE_\lambda$. Then its mathematical expectation value for a sate $\psi$ 
 can be expressed as:
\[
\overline{A}_\psi = \int_{-\infty}^{+\infty}\lambda dp_\psi (\lambda),
\]
where $dp_\psi (\lambda) = d\langle E_\lambda\psi, \psi\rangle$ is the probability distribution for the corresponding spectral decomposition and quantum state. Therefore for the fixed $\psi$ quantum observable can be regarded as a classical random variable with the probability distribution
$p_\psi (O) = \int_O d\langle E_\lambda\psi, \psi\rangle, O\subset  \mathbb{R}$.

Consider the following problem. Let the observable $A$ depend on some classical random variable 
$\omega: A = A(\omega)$ corresponding to the case when angle values for the prisms positions 
cannot be set without an error during experiments. In this case the spectral decomposition 
$A(\omega) = \int_{-\infty}^{+\infty}\lambda dE_\lambda(\omega)$ and the density distribution function 
$dp_\psi (\lambda|\omega) = d\langle E_\lambda(\omega)\psi, \psi\rangle$ also depends on 
this random variable. For every fixed $\omega$ also 
$\int_{-\infty}^{+\infty}dp_\psi (\lambda|\omega) = 1$ condition holds.

Let the random variable $\omega$ be described using the Kolmogorov probability space $(\Omega, \mathcal{F}, P)$,
 where $\Omega$ is the set of elementary events, $\mathcal{F}$ is the $\sigma$-algebra of events and $P$ is the probability measure. The mathematical expectation value of $A(\omega)$ takes the following form:
\[
\overline{A}_\psi(\omega) = \int_\Omega\left[\int_{-\infty}^{+\infty}\lambda dp_\psi (\lambda|\omega)\right] dP(\omega) = \int_\Omega\langle A(\omega)\psi, \psi\rangle dP(\omega) = \mathbf{\tilde{E}}[ \langle A(\omega)\psi, \psi\rangle],
\]
where $\mathbf{\tilde{E}}[\cdot]$ is the classical mathematical expectation.

In a similar way we obtained an expression for the dispersion $A(\omega)$:
\[
\sigma^2(A) = \mathbf{\tilde{E}}[ \langle A^2(\omega)\psi, \psi\rangle - \langle A(\omega)\psi, \psi\rangle^2].
\]

\section{The model with uniform random fluctuations of four angles of polarization beam splitters}

Consider a model in which the value of each angle in the experiment is uniformly distributed at the section around the desired value. In this case 
mathematical expectation and dispersion values become: 
\[
J_\mathcal{B} = \frac{1}{16\delta^4}\int_{-\delta}^\delta\int_{-\delta}^\delta\int_{-\delta}^\delta\int_{-\delta}^\delta \langle \mathcal{B}(x_1, x_2, x_3, x_4)\psi, \psi \rangle dx_1dx_2dx_3dx_4,
\] 
\[
\sigma^2 = \frac{1}{16\delta^4}\int_{-\delta}^\delta\int\int_{-\delta}^\delta\int_{-\delta}^\delta (\langle \mathcal{B}^2\psi, \psi \rangle - \langle \mathcal{B}\psi, \psi \rangle^2) dx_1dx_2dx_3dx_4.
\]

\begin{figure}[h]
\includegraphics[scale=0.3]{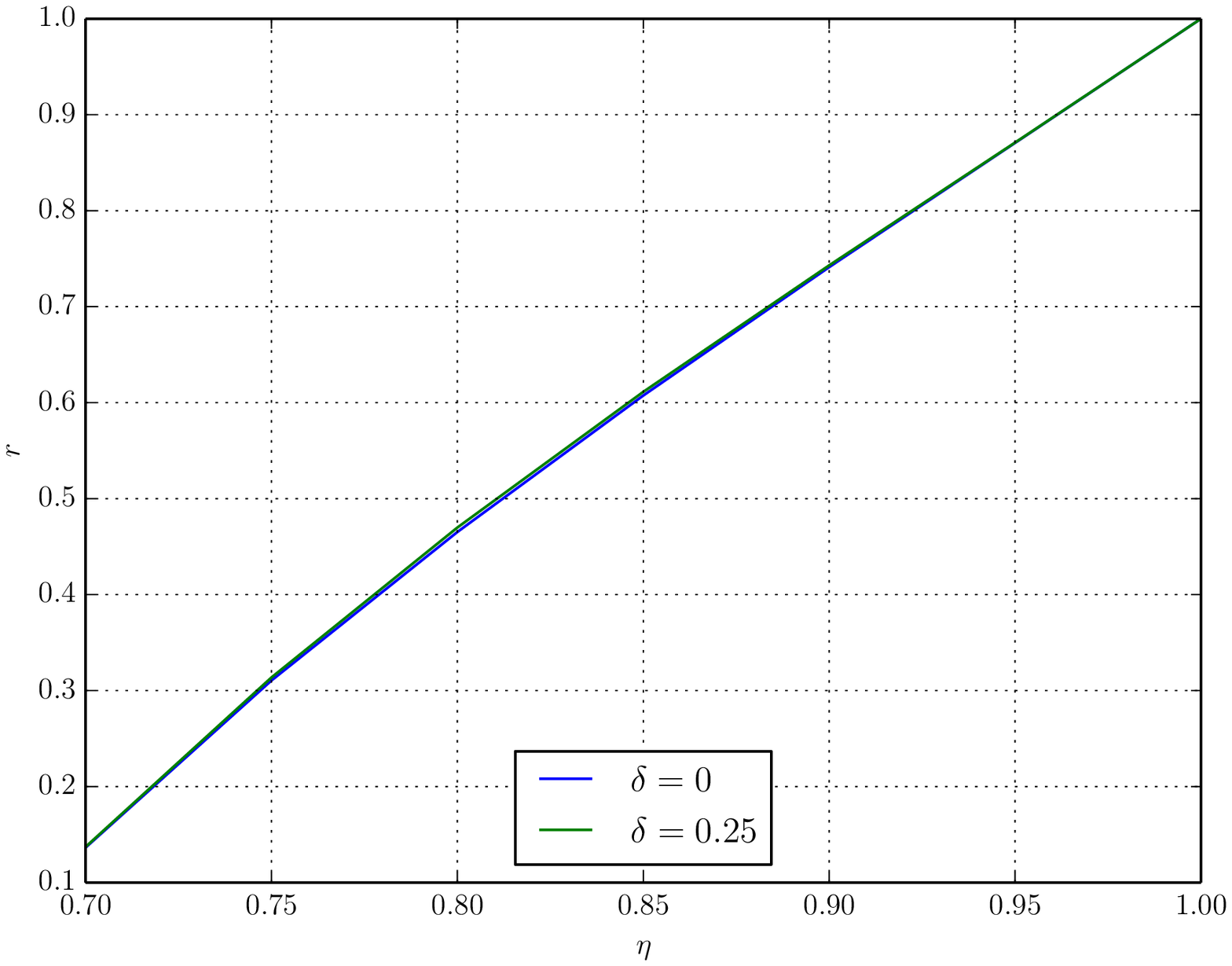}
\includegraphics[scale=0.3]{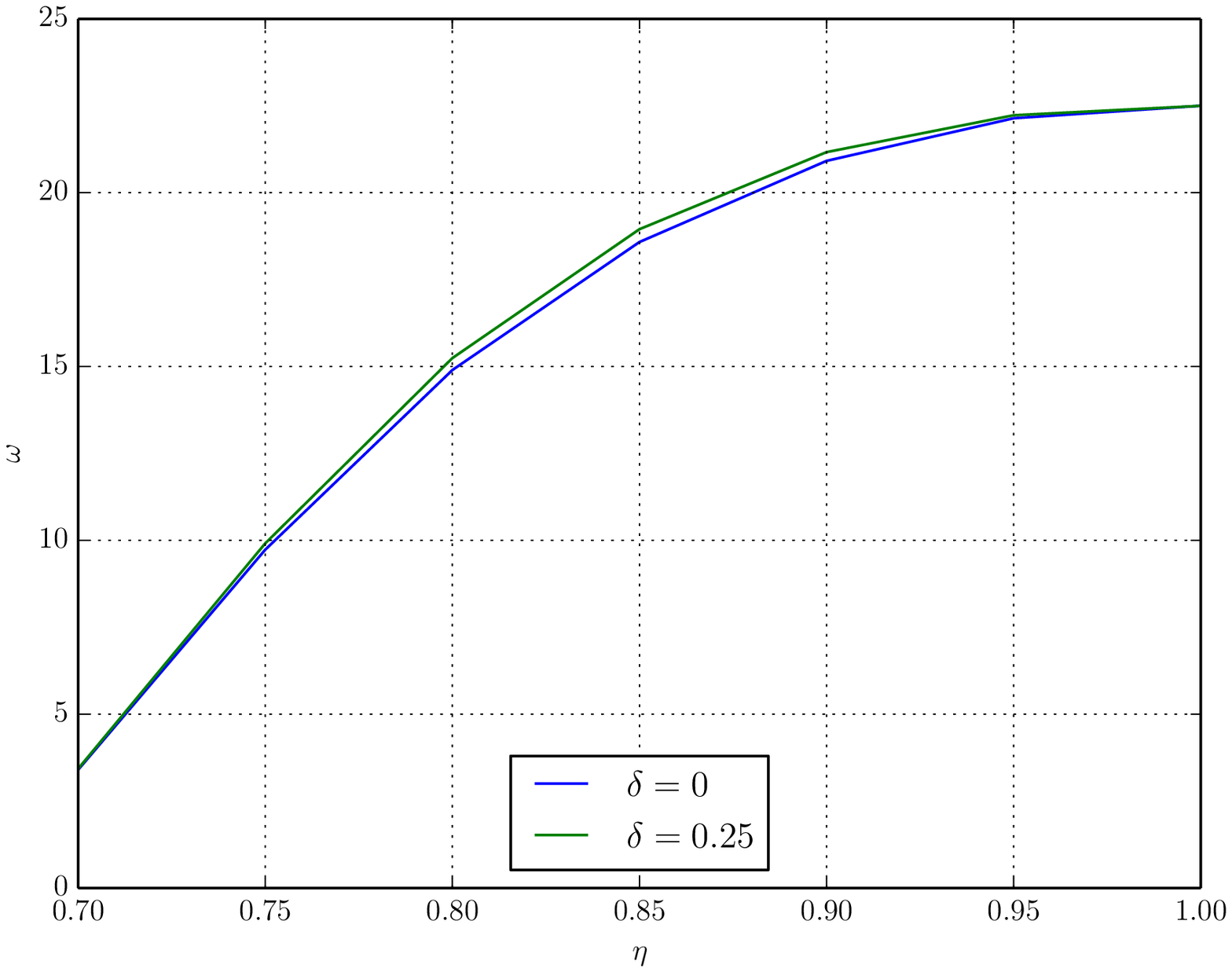}
\caption{Optimized $r$ and $\omega$ values  (depending on $\eta$) for various values of angles}
\label{fig:psi_4ang}
\end{figure}

\begin{figure}[t]
 \includegraphics[scale=0.3]{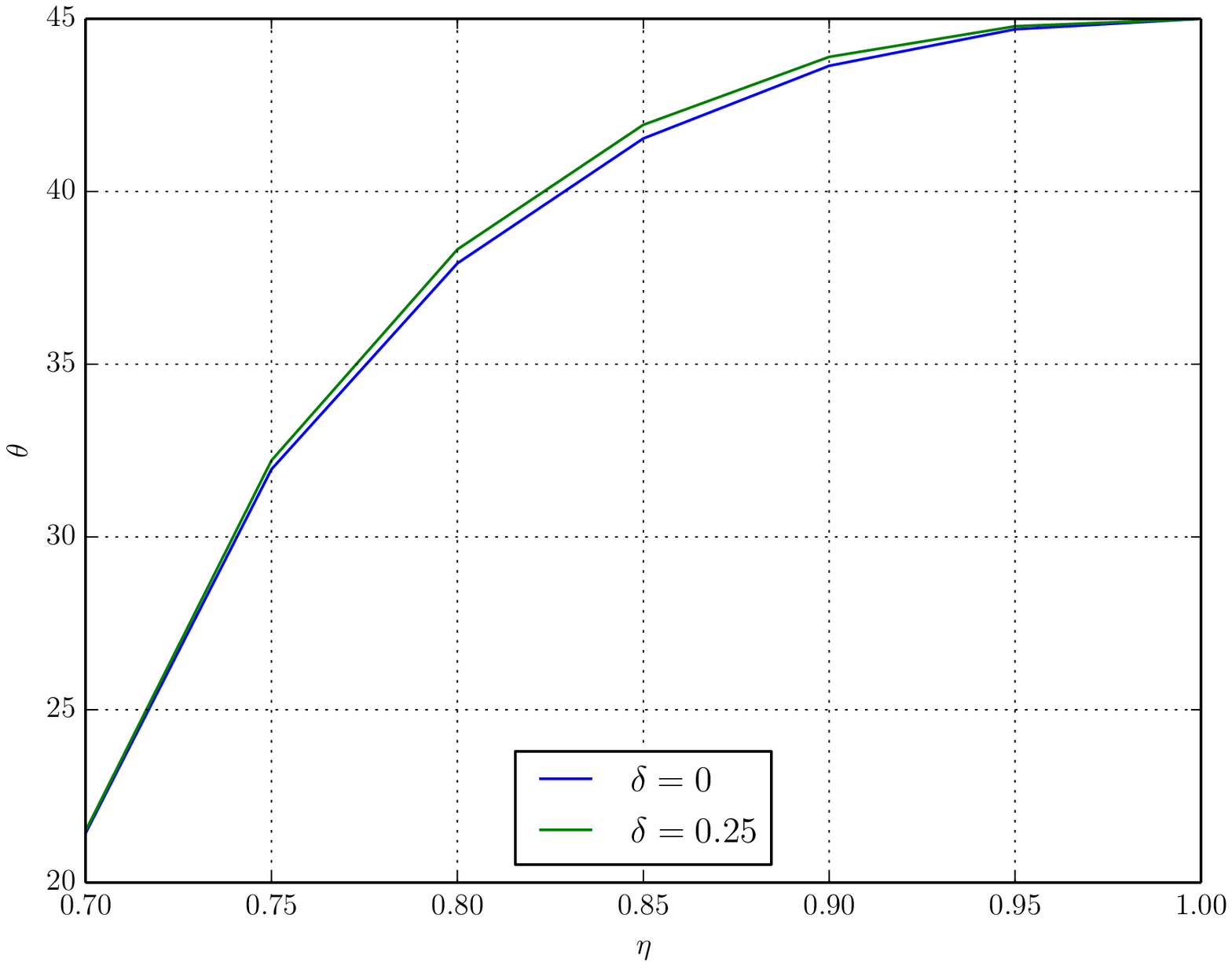} 
\includegraphics[scale=0.3]{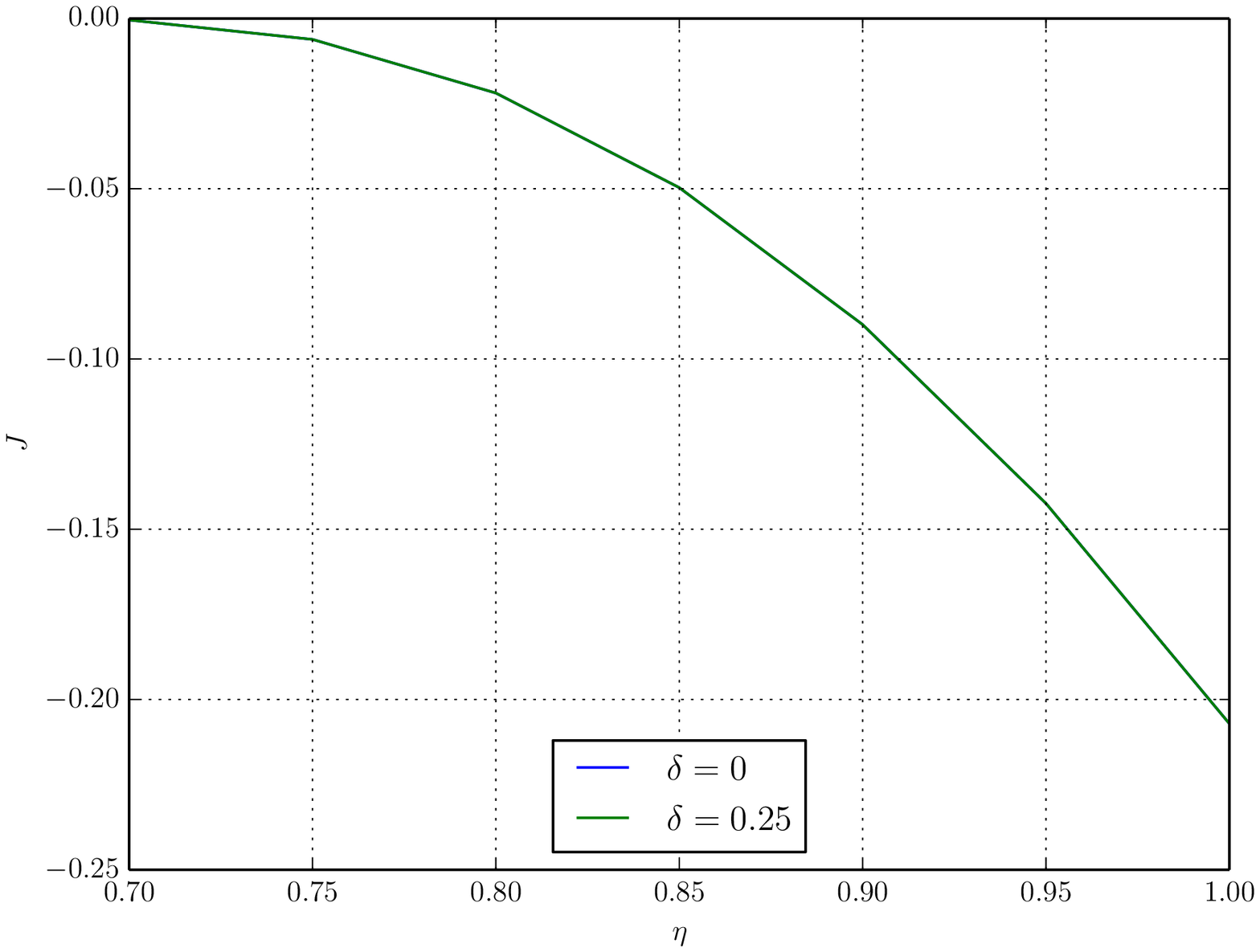}
\caption{Optimized $\theta$ and $J_{\mathcal{B}}/N$ values (depending on $\eta$) for various values of angles}
\label{fig:J_4ang}
\end{figure}

\begin{table}
\small
\begin{tabular}{|c|c|c|c|c|c|c|c|c|}
\hline
$\eta$ & $\delta,^\circ$ & $r$ & $\omega,^\circ$ & $\theta,^\circ$ & $J$ & $J_{\delta}$ & $\sigma_{\delta}$ & $K = J_{\delta}/\sigma_{\delta}$
\\\hline
\multirow{3}{*}{0.7} & \multirow{3}{*}{0.25} & 0.136389 & 3.40081 & 21.4266 & -0.000453562 & -0.000444565 & 0.00241554 & -0.184044\\\cline{3-9}
 &  & 0.136389 & 3.40081 & 21.4266 & -0.000453562 & -0.000444565 & 0.00241554 & -0.184044\\\cline{3-9}
 &  & 0.137124 & 3.42997 & 21.496 & -0.000453514 & -0.000444515 & 0.00241503 & -0.184062\\\hline
 \multirow{3}{*}{0.75} & \multirow{3}{*}{0.25} & 0.310518 & 9.73143 & 31.9603 & -0.00615095 & -0.00614082 & 0.00248895 & -2.46724\\\cline{3-9}
 &  & 0.310518 & 9.73143 & 31.9603 & -0.00615095 & -0.00614082 & 0.00248895 & -2.46724\\\cline{3-9}
 &  & 0.313658 & 9.91344 & 32.2158 & -0.00614786 & -0.00613773 & 0.00248642 & -2.4685\\\hline
\multirow{3}{*}{0.8} & \multirow{3}{*}{0.25} & 0.465228 & 14.8979 & 37.9215 & -0.02191 & -0.0218985 & 0.002596 & -8.43546\\\cline{3-9}
 &  & 0.465228 & 14.8979 & 37.9215 & -0.02191 & -0.0218985 & 0.002596 & -8.43546\\\cline{3-9}
 &  & 0.469841 & 15.2419 & 38.3231 & -0.0218953 & -0.0218838 & 0.00259252 & -8.44116\\\hline
\multirow{3}{*}{0.85} & \multirow{3}{*}{0.25} & 0.607424 & 18.5808 & 41.5341 & -0.0496902 & -0.0496772 & 0.00275221 & -18.0499\\\cline{3-9}
 &  & 0.607424 & 18.5808 & 41.5341 & -0.0496902 & -0.0496772 & 0.00275221 & -18.0499\\\cline{3-9}
 &  & 0.61123 & 18.9498 & 41.9292 & -0.0496699 & -0.0496569 & 0.00274992 & -18.0576\\\hline
\multirow{3}{*}{0.9} & \multirow{3}{*}{0.25} & 0.741202 & 20.9153 & 43.6381 & -0.0899078 & -0.0898932 & 0.00296469 & -30.3213\\\cline{3-9}
 &  & 0.741202 & 20.9153 & 43.6381 & -0.0899078 & -0.0898932 & 0.00296469 & -30.3213\\\cline{3-9}
 &  & 0.743038 & 21.167 & 43.8961 & -0.0898969 & -0.0898824 & 0.00296395 & -30.3252\\\hline
\multirow{3}{*}{0.95} & \multirow{3}{*}{0.25} & 0.87067 & 22.141 & 44.6958 & -0.142436 & -0.14242 & 0.00323493 & -44.0258\\\cline{3-9}
 &  & 0.87067 & 22.141 & 44.6958 & -0.142436 & -0.14242 & 0.00323493 & -44.0258\\\cline{3-9}
 &  & 0.871004 & 22.2272 & 44.7823 & -0.142435 & -0.142419 & 0.00323486 & -44.0263\\\hline
\multirow{3}{*}{1.0} & \multirow{3}{*}{0.25} & 0.999997 & 22.5 & 45 & -0.207107 & -0.207089 & 0.0035626 & -58.1286\\\cline{3-9}
 &  & 0.999997 & 22.5 & 45 & -0.207107 & -0.207089 & 0.0035626 & -58.1286\\\cline{3-9}
 &  & 0.999999 & 22.4981 & 44.998 & -0.207107 & -0.207089 & 0.0035626 & -58.1286\\\hline
 \end{tabular}
\caption{Optimized parameters' values for the error in four angles separately in case of 
$\delta = 0.25^\circ$}
\label{tab:different_deltas}
\end{table}

Results of the performed optimization are shown at Fig.
\ref{fig:psi_4ang}, \ref{fig:J_4ang}  and in  
Table~\ref{tab:different_deltas}, its rows are also grouped in triads. It follows from the graphs that 
the addition of random fluctuations of the angles almost do not change optimal parameters. Therefore we can suggest 
that angles' values control can be reduced.

\clearpage

\section{Conclusion}

In this paper we analyzed the Eberhard inequality \cite{Eberhard}. 
(This inequality obtained in 1993 was practically forgotten, experimentalists and theoreticians were interested mainly 
in the CHSH-inequality.)
Our goal was to find angles of 
polarization beam splitters and a quantum state (which is entangled, but not maximally entangled) 
that allow to violate the inequality as much as possible.
Required parameters were found using optimization procedure based on the Nelder-Mead optimization method \cite{NM}. 
We considered two models, one due to Eberhard \cite{Eberhard}, and another used for the Vienna-13 experiment, see 
Gustina et al.  \cite{Zeilinger} and Kofler et al. \cite{Zeilinger1}. In the first case we obtained values consistent 
with the values from the article \cite{Eberhard}. Note that the model of Eberhard describes only 
the case of  equal detector efficiencies. However, in real experiments, detectors' efficiencies may differ essentially. 
Therefore it was important to perform the study similar to \cite{Eberhard} for detectors of different efficiencies. And such a study 
was done. In the second case (Vienna-13) we obtained values of parameters which differ slightly from the values used
for the Vienna-13 experiment  \cite{Zeilinger}, \cite{Zeilinger1}. Our optimal values of the angles and the state parameters 
provide a possibility to obtain stronger violation of the Eberhard inequality than in  \cite{Zeilinger}, \cite{Zeilinger1}.
We remark that the model of Kofler et al. \cite{Zeilinger1} is asymmetric with respect to detectors' efficiencies. We explored
this feature of the model and found (that is curious) that experimentalists from Vienna would be able to obtain a stronger violation
of the Eberhard inequality simply by permutation of the detectors which they used for the experiment. 

In both aforementioned models it was assumed that in real experiments the optimal values of the angles of polarization beam splitters
(obtained as the results of optimization) 
 can be fixed in the perfect accordance with the theoretical prediction. Although this assumption is justified up to a high degree,
 in real experiments there are always present errors in fixing of these angles. Such random errors have to be taken into account.
This is an important part of this paper. We performed the corresponding theoretical modeling completed with numerical simulation. 
In this model
the magnitude of the possible spread of experimental data which can be expressed using the reciprocal of the {\it coefficient of 
variation} $\sigma_J / J $ (also known as signal/noise ratio) was studied.
The obtained parameters differ from the results of Eberhard \cite{Eberhard} and Gustina et al. \cite{Zeilinger} 
and Kofler et al. \cite{Zeilinger1}. The simulation results can be interesting for experimenters 
as they allow to weaken control over the precision of orientation of the axes of polarization beam splitters. 

The obtained results allow us to expect that in the experimental test of the Bell type inequality in the Eberhard form with 
the optimal values of physical parameters from this paper, the inequality will be significantly violated even 
for different detectors' efficiencies and inaccuracy in the installation angles and without the assumption of the purity of the 
initial state. We hope that our study may be useful for experimentalists trying to perform a loophole free Bell test, i.e., trying 
to combine closing of the detection loophole with closing of the locality loophole.

\section*{Acknowledgments}
This work was partially supported by  a visiting fellowship of A. Khrennikov (May-June, 2014) to  
Institute for Quantum Optics and Quantum Information, Austrian Academy of Sciences. The visit of P. Titova
to Linnaeus University was supported by a student fellowship of Moscow University of Electronic Technology.
One of the authors (A. Khrennikov)  would like to thank I. Basieva, M. Gustina, J. Kofler, S. Ramelow, R. Ursin, and B. Wittmann for 
fruitful discussions about the Vienna-13 experiment and knowledge transfer and A. Zeilinger for numerous discussions on quantum 
foundations, including the problem of ``technicalities'' related to a loophole free Bell test.

\end{document}